\documentstyle[twoside,fleqn,espcrc2,epsf]{article}

\title{\hfill \normalsize TUM-T31-71/94 \\[-2truemm]
\hfill \normalsize hep-ph/9408401 \\[-2truemm]
\hfill \normalsize August 1994 \\
The determination of $|V_{cb}|$ from semileptonic inclusive
       decay rates}

\author{Ulrich Nierste\address{Physik--Department T31, Technische Universit\"at
        M\"unchen, 85747 Garching,
        Germany}\thanks{supported by BMFT under grant no.\ 06-TM-732.
Talk given at the QCD 1994 conference in Montpellier, France,
7th to 13th July 1994.} }

\begin{document}

\begin{abstract}
We report on the determination of $|V_{cb}|$ from the comparison of the
semileptonic inclusive decay rates of  B-- and D--mesons using the heavy
quark symmetry of QCD.
While the renormalization scale ambiguity
does not allow for a  reliable estimate of the
quark masses,  it almost cancels in the prediction for $|V_{cb}|$, which reads
\begin{center}
   $|V_{cb}|(\tau_B/1.49\,{\rm ps})^{1/2} = 0.036\pm 0.005$,
\end{center}
where the error stems from the uncertainty in the kinetic energy
of the heavy quark inside the meson, in the experimental branching
ratios, in QCD input parameters, and from scale uncertainties.
\end{abstract}

\maketitle

\section{Introduction}
The following talk is based on research done in collaboration with
Patricia Ball \cite{bn}.

In the recent years the study of HQET, the effective theory of QCD
expanded in inverse powers of the heavy quark mass $m_Q$,
has considerably
enlarged our understanding of low--energy  QCD.
%
Bigi, Uraltsev and Vainshtein \cite{buv} have started to apply
HQET to inclusive processes and have been succeeded by many
authors.
Two important
statements about inclusive decays could be obtained: first, in leading
order they are essentially free quark decays, and second, the leading
corrections to the free quark decay are of order $1/m_Q^2$.
These results stimulated new determinations of the quark masses $m_c$
and $m_b$ and the CKM matrix element $|V_{cb}|$ from the experimental
measurements of the semileptonic branching ratios $B(D\to Xe\nu)$
and $B(B\to X_c e \nu)$
\cite{ls,bu,ln}.
Now  the determination of $|V_{cb}|$ from {\em exclusive} decay rates
requires the analysis of the few experimental data points near the endpoint
of the lepton spectrum and moreover needs some
model--dependent input on a formfactor. The {\em inclusive} decay rates
to order $1/m_Q^2$, however,  only  involve two real parameters $\lambda_1$
and $\lambda_2$, of which the latter is well known from the
$B^\ast - B$ mass splitting. On the other hand they are proportional to
the  fifth powers of the poorly known quark masses, which are moreover
renormalization--scheme dependent quantities. These points will be
discussed in the following section, where the analytic expression
for the inclusive decay rate will be introduced. Section 3 will
in detail describe the phenomenological analysis.

\section{The inclusive decay rate}
When physical observables are calculated with the help of HQET,
at first some QCD Green function is matched to its counterpart in
HQET.
In the case of the semileptonic
inclusive decay rate this is the self energy
$\Sigma$  of the decaying heavy quark with a lepton pair and a quark
in the intermediate state:
\begin{eqnarray}
\lefteqn{\Sigma _{\rm QCD} ( m_Q,\mu _Q ) \;  = } \nonumber \\
&& \!\!\!\!\!\!
 C_1 ( m_Q , \mu _Q )
       \left[ \Sigma_{\rm HQET}^{(0)}
           + \frac{1}{2 m_{Q,{\rm pole}}^2 }
       \Sigma_{\rm HQET}^{(1)}
  \right] \nonumber \\
&& \!\!\!\!\!\! +
C_2 ( m_Q , \mu _Q )
            \frac{1}{2 m_{Q,{\rm pole}}^2 }
       \Sigma_{\rm HQET}^{(2)}
    + O \! \left(\frac{1}{m_Q^3}\right) , \label{match}
\end{eqnarray}
whose imaginary part
is related to the desired rate via the optical theorem.
In (\ref{match})
$C_1$ and $C_2$ are Wilson coefficients and the HQET matrix elements
read in terms of the heavy quark field $h$:
\begin{eqnarray}
\Sigma _{\rm HQET} ^{(0)}\!\!\! &=& \!\!\! \frac{1}{2 M_Q}
       \langle {\cal M} | \bar{h} h | {\cal M} \rangle
              =  1 \nonumber , \\
\Sigma _{\rm HQET} ^{(1)} \!\!\! &=& \!\!\! \frac{1}{2 M_Q}
       \langle {\cal M} | \bar{h} (iD )^2 h | { \cal M }\rangle
              =  \lambda _1 \nonumber , \\
\Sigma _{\rm HQET} ^{(2)}\!\!\! &=& \!\!\! \frac{1}{6 M_Q}
       \langle {\cal M} | \bar{h} \frac{g}{2}  \sigma _{\mu \nu }
                 F^{\mu \nu }   h | { \cal M }  \rangle
              =  \lambda _2 (\mu _Q) , \nonumber
\end{eqnarray}
where $\cal M$ is the heavy meson  with mass $M_Q$. Heavy quark symmetry
dictates that the $\lambda _i$'s  are the same for
${\cal M}=B$ and ${\cal M}=D$.
The matching scale $\mu _Q$ in (\ref{match}) must be of order $m_Q$,
where both perturbative QCD and HQET are valid.
As indicated in (\ref{match}) the expansion parameter in the HQET
matrix elements is the QCD pole mass. The Wilson coefficients $C_1,
C_2 \propto m_Q ^5$, however,
contain the mass parameter of the renormalization scheme
chosen to calculate the QCD Green function on the left hand side of
(\ref{match}).
In view of the fact that the Wilson coefficients
contain the short distance physics from scales larger than $\mu _Q$, while
the interaction  from lower scales is contained in the matrix elements,
we have used a short distance mass $m_Q^{\overline{\rm MS}}$
evaluated at the matching scale $\mu _Q$
in the $C_i$'s. Clearly, the fifth power of $m_Q$ causes a sizeable
dependence of the decay rate on the renormalization scheme used
to define $m_Q$.
Recently Beneke and  Braun  and Bigi et al.\ \cite{bbb} have  found
that  the perturbation series defining the pole mass $m_{\rm pole}$
suffers from an extra IR--renormalon  imposing an ambiguity of order
$\Lambda _{\rm QCD}$ onto $m_{\rm pole}$. When the inclusive decay rate
is expressed in terms of $m_{\rm pole}$, the perturbation series
multiplying $m_{\rm pole}^5$ exhibits the same renormalon ambiguity,
which, however,  vanishes, when the rate is expressed
in terms of some short distance mass \cite{bbz}.

Taking  the absorptive part of (\ref{match}) results in
\begin{eqnarray}
\lefteqn{
\Gamma( {\cal M} \to X \ell \nu) =
     \frac{ G_F ^2 (m^R_Q)^5}{192\pi^3} |V_{\rm CKM}|^2
       \cdot } \nonumber \\
 \!\!\!\! &&
\left[ \left\{ 1-\frac{2}{3}\,\frac{\alpha_s^R (\mu_Q)}{\pi}\,
g^R\!\left(  x^R \right) + \frac{\lambda_1}{2m_{Q}^2} \right\}
f_1\!\left(x^R \right) \right. \nonumber \\
\!\!\!\! &&
\left. {}  - \frac{9 \lambda_2}{2 m_{Q} ^2}\,
f_2\!\left( x^R \right) + {\cal O}\left( \frac{1}{m_Q^3},
(\alpha_s^R)^2, \frac{\alpha_s^R}{m_Q } \right) \right].
\label{gamma}
\end{eqnarray}
Here $R$ marks the renormalization scheme dependent quantities.
In the following section we exploit (\ref{gamma}) for
$({\cal M} ,  m^R_Q ,V_{\rm CKM}, x^R    )=
  ( B , m_b^{\overline{\rm MS}} (\mu_b) ,V_{cb},
  m_c^{\overline{\rm MS}}(\mu _b) /m_b^{\overline{\rm MS}} (\mu_b)  ) $
and
$ = ( D , m_c^{\overline{\rm MS}} (\mu_c) ,V_{cs},
  m_s^{\overline{\rm MS}}(\mu_c)/m_c^{\overline{\rm MS}}(\mu_c)  ) $.
The matching scales $\mu_b \approx m_b $  and $\mu_c \approx m_c $
will be varied to judge the renormalization scale dependence.
The analytic expressions for $g(x)$ and the phase space factors $f_i(x)$
can be found in \cite{bn}.

\section{The determination of $|V_{cb}|$}
The input of our phenomenological analysis is similar to \cite{ls,bu,ln}.
It consists of four steps:

{\bf Step 1:}
Extract $m_c^{\overline{\rm MS}}(m_c)$ from
the experimental result for
$\Gamma (D \to X e \bar{\nu})$ as given in (\ref{gamma})
as a function of
$\mu_c, \lambda_1, m_s(1 \mbox{GeV})$ and $ \Lambda_{\overline{\rm MS}}  $.

{\bf Step 2:}
Get $m_b^{\overline{\rm MS}}(m_b)$  from
 $m_c^{\overline{\rm MS}}(m_c)$ via the heavy quark symmetry,
which relates pole masses:
\begin{eqnarray}
m_b^{\overline{\rm MS}}(m_b) & = &   m_c^{\overline{\rm MS}}(m_c) +
\Sigma_b^{(1)} -\Sigma_c^{(1)}  +
m_B  \nonumber \\
&& - m_D + \frac{\lambda_1+3 \lambda_2}{2 m_b} -
  \frac{\lambda_1+3 \lambda_2}{2 m_c},
\end{eqnarray}
where $\Sigma_q^{(1)}$ is the one--loop QCD quark self energy.

{\bf Step 3:}
Insert $m_b^{\overline{\rm MS}}(m_b)$ into  (\ref{gamma}), but now  for
$\Gamma (B \to X_c e \bar{\nu})$, to find $|V_{cb}|$.

{\bf Step 4:}
Vary the two matching scales $\mu_c$ and $\mu_b$ to estimate the
renormalization scale dependence and also the physical paramaters
$\lambda_1, m_s^{\overline{\rm MS}}(1 \mbox{GeV}) \ldots$ to judge
the total error of the theoretical prediction.

The one--loop expression for the decay rate (\ref{gamma})
exhibits a large scale dependence, especially for the D--decay.
This fact obscurs the determination of the quark masses
as displayed in fig.\ 1., for which we find
\begin{eqnarray}
m_c^{\overline{\rm MS}}(m_c) &=& (1.35\pm 0.20)\,\mbox{GeV} , \nonumber
\\
m_b^{\overline{\rm MS}}(m_b) &=& (4.6\pm 0.3 )\,\mbox{GeV}. \label{mass}
\end{eqnarray}
\begin{figure}[htb]
\centerline{
\epsfxsize=0.35 \textwidth
\epsfbox{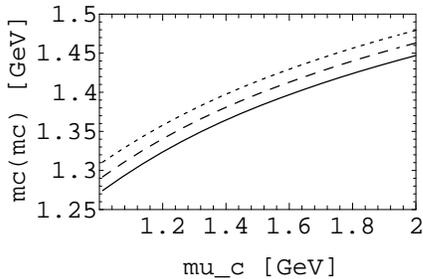}}
\vspace{-5ex}
\caption[]{\small $m_c(m_c)$ vs.\ the renormalization scale
$\mu_c$ for  the
branching ratio $B(D\to Xe\nu)= 0.172$,
$m_s(1\,\mbox{GeV}) = 0.2\,\mbox{GeV}$,
$\Lambda^{(4)}_{\overline{\rm MS}} = 300\,\mbox{MeV}$.
Solid, \mbox{long--,} short--dashed line:
$\lambda_1 = 0\,\mbox{GeV}^2 ,
-0.35,-0.7 \,\mbox{GeV}^2$. \vspace{-2ex}  }
\end{figure}
This shows that one has to calculate higher orders in (\ref{gamma})
in order to reduce the scale dependence,
if one wants to extract quark masses from the inclusive decay rates.
In the determination of $|V_{cb}|$ the scale dependence, however,
reduces drastically as displayed in fig.\ 2.
\begin{figure*}[tb]
\centerline{
\epsfxsize=0.7 \textwidth
\epsfbox{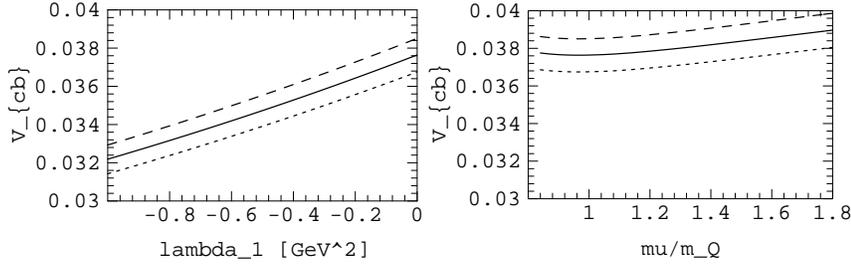}}
\vspace{-5ex}
\caption[]{\small \underline{Left}:
$|V_{cb}|(\tau_B/1.49\,\mbox{ps})^{1/2}$ vs.\
$\lambda_1$ for $\mu_Q=m_Q $.
\underline{Right}:
$|V_{cb}|(\tau_B/1.49\,\mbox{ps})^{1/2}$ vs.\
$\mu_Q/m_Q$ for $\lambda_1 = 0 $.
Solid line:
$B(D\to Xe\nu)= 0.172$, $m_s(1\,\mbox{GeV}) =
0.2\,\mbox{GeV}$ and $\Lambda^{(4)}_{\overline{\rm MS}}
= 300\,\mbox{MeV}$ and
$B(B\to X_c e\nu)= 0.107$.
The dashed lines correspond to the the experimental error in
$B(B\to X_ce\nu)$:
$B=0.102$ (short),
$B=0.112$ (long).
}
\end{figure*}
Here we have taken $\mu_c/m_c=\mu_b/m_b$, because by varying the
scale we want to estimate the importance of the yet uncalculated
higher order terms in the perturbation series in (\ref{gamma}), as
the scale dependence vanishes order by order in perturbation theory.
These uncalculated terms are of course the same function
of $\mu_Q/m_Q$ for
$\Gamma (B \to X_c e \bar{\nu})$  and for
$\Gamma (D \to X  e \bar{\nu})$ apart from the small effect that one
has five active flavours in the former rate and four in the latter.
Suppose the $O(\alpha_s^2)$--corrections
enhance $\Gamma$ in
(\ref{gamma}):
Then both the error and the central value for $m_c$ and
$m_b$ in (\ref{mass}) obtained in step 1 and 2 will be lower,
but in step 3 this  lower value  for $m_b^5$ will  multiply
a larger radiative correction to $\Gamma $ in (\ref{gamma}),
thereby stabilizing the prediction for $ | V_{cb} |$.

Finally the largest uncertainty in $|V_{cb}|$ originates from
$\lambda_1$ as shown in fig.\ 2. We have varied $\lambda_1$
between -0.7 GeV${}^2$ \cite{bb} and  0 \cite{n} and find:
\begin{eqnarray}
|V_{cb}|\left(\frac{\tau_B}{1.49\,\mbox{ps}}\right)^{1/2} \!\!\! \!
&=& \!\!\! \! 0.036 \pm 0.005   \nonumber .
\end{eqnarray}
This has to be compared with
\begin{eqnarray}
|V_{cb}|\left(\frac{\tau_B}{1.49\,\mbox{ps}}\right)^{1/2} &= &
(0.046\pm 0.008)\quad \mbox{\protect{\cite{ls}},}\nonumber\\
|V_{cb}|\left(\frac{\tau_B}{1.49\,\mbox{ps}}\right)^{1/2} & \approx &
0.042\quad\mbox{\protect{\cite{bu}}.}\nonumber
\end{eqnarray}
We remark that in \cite{ls} the renormalization scale has been
varied down to $\mu_c \approx 0.5$GeV, which is too low to trust into
perturbative QCD.
Recent exclusive measurements (CLEO) gave \cite{c}:
\begin{eqnarray}
|V_{cb}| &=& 0.0362 \pm 0.0053 . \nonumber
\end{eqnarray}

\end{document}